\documentclass[twocolumn,prc,superscriptaddress,unsortedaddress,showpacs,preprintnumbers,amsmath,amssymb]{revtex4}

\usepackage{graphicx}
\usepackage{dcolumn}
\usepackage{bm}
\usepackage{CJK}
\usepackage{CJKulem}
\usepackage{txfonts}

\def\beq{\begin{equation}}
\def\eeq{\end{equation}}
\def\bea{\begin{eqnarray}}
\def\eea{\end{eqnarray}}

\def\fun#1#2{\lower3.6pt\vbox{\baselineskip0pt\lineskip.9pt
  \ialign{$\mathsurround=0pt#1\hfil##\hfil$\crcr#2\crcr\sim\crcr}}}

\begin{document}
\begin{CJK*} {GBK} {song}
\preprint{}

\title{Magnetization of neutron star matter}

\author{Jianmin Dong}\affiliation{Institute of Modern Physics, Chinese Academy of
Sciences, Lanzhou 730000, China}
\author{Wei Zuo}\email[ ]{zuowei@impcas.ac.cn}
\affiliation{Institute of Modern Physics, Chinese Academy of
Sciences, Lanzhou 730000, China}
\author{Jianzhong Gu}
\affiliation{China Institute of Atomic Energy, P. O. Box 275(10),
Beijing 102413, China}

\date{\today}

\begin{abstract}
\noindent The magnetization of neutron star matter in magnetic
fields is studied by employing the FSUGold interaction. It is found
that the magnetic susceptibilities of the charged particles (proton,
electron and muon) can be larger than that of neutron. The effects
of the anomalous magnetic moments (AMM) of each component on the
magnetic susceptibility are examined in detail. It is found that the
proton and electron AMM affect their respective magnetic
susceptibility evidently in strong magnetic fields. In addition,
they are the protons instead of the electrons that contribute most
significantly to the magnetization of the neutron star matter in a
relative weak magnetic field, and the induced magnetic field due to
the magnetization can be appear to be very large. Finally, the
effect of the density-dependent symmetry energy on the magnetization
is discussed.
\end{abstract}


\maketitle
\section{Introduction}\label{intro}\noindent

Neutron stars in the universe tend to contain matter of supranuclear
density in their interiors, with typical mass $M\sim 1.4M_{\odot }$
and radii $R\sim 10$km. As one class of compact objects, neutron
stars have been arousing tremendous interest amongst scientists
because of many novel features. One of the features of neutron stars
is their strong magnetic field that could be the largest one
observed in nature. The typical magnitudes of a surface magnetic
fields are as large as $10^{11}-10^{13}$ G \cite{HPY}. It is
currently assumed that the soft gamma repeaters (SGR) and anomalous
X-ray pulsars, candidates for the magnetars, have a strong surface
magnetic fields up to $10^{14}-10^{15}$ G \cite{MAS}. The magnetic
field in the interior could be as large as $10^{18}$ G according to
the scalar virial theorem \cite{DL}. It is interesting that the
strong magnetic fields were also created in heavy-ion collisions
\cite{Au1,Au2}, which may help us to understand the response of the
dense matter under the presence of strong magnetic fields.

Over the past decades, many works have been dedicated to the effects
of the magnetic field on neutron star properties, such as the
equation of states \cite{EoS1}, neutron star structure \cite{NS0},
transport properties and the cooling or heating of magnetized stars
\cite{Cool}. An unclear but interesting problem is the origin of
such strong magnetic field. A simple analysis showed that a weak
magnetic field in a progenitor star could be amplified during the
gravitational collapse due to magnetic flux conservation. However,
it can not explain the very strong surface magnetic field in
magnetars \cite{TTT}. Another explanation called the
magnetohydrodynamic dynamo mechanism based on the rapidly rotating
plasma of a protoneutron star \cite{TDM} which is generally accepted
as the standard explanation for the origin of the magnetar's large
magnetic fields, is unable to explain all the features of the
supernova remnants surrounding these objects \cite{JV,XRX}. An
interesting mechanism being suggested for the origin is the possible
existence of a phase transition to a ferromagnetic state, namely
spontaneous magnetization. Such argument has been investigated
widely within various theoretical approaches (without the background
magnetic field) \cite{AAAA}, but the results are still divergent.
Even some authors showed that a possibility that a strong magnetic
field is produced by color ferromagnetic quark matter in neutron
stars \cite{QCD}. Astronomical observations found that the SGR
1806-20 emitted a giant flare on 27 December 2004 with the total
flare energy by $2\times10^{46}$ erg and the energy release probably
occurred during a catastrophic reconfiguration of the neutron star's
magnetic field since the emitted energy significantly exceeds the
rotational energy loss in the same period \cite{DMP}. These
phenomena are perhaps related to the magnetization of the neutron
star matter. In addition, the anisotropic pressure is related to the
magnetization for the magnetized matter \cite{GM5}. Therefore, the
magnetization is an important physical quantity for neutron stars.

Some calculations have been performed for the magnetization of
nuclear matter or pure neutron matter in magnetic fields
\cite{GM5,GM4,GM6}. Seldom calculations were carried out for the
$\beta$-stable matter. In Ref. \cite{GM3}, the magnetization of the
$\beta$-stable matter was studied and it is shown that the
magnetization never appears to become very large. However, this
conclusion could be revised according to our calculations, as shown
later. Because of the small mass and hence the small magneton, the
magnetization of electrons may be important compared with that of
neutron. Therefore, in the present study, the magnetization of the
$\beta$-stable neutron star matter, which consists of protons,
neutrons, electrons and muons, will be investigated. Not only the
AMM of nucleons but also the one of leptons are included here. The
main purposes of the this study are as follows. Firstly, the
contribution of each component as well as the effect of the
anomalous magnetic moments (AMM) will be analyzed in detail.
Secondly, we further explore whether the strong magnetic fields of
the neutron stars originate from the highly degenerate relativistic
electron gas. Finally, the symmetry energy effects on the
magnetization will be presented.

This work is organized as follows. In Sec. II, a brief introduction
of the relativistic mean field approach is presented. The
magnetization of each component of the neutron star matter, along
with the effects of the AMM and the symmetry energy, are analyzed in
detail in Sec. III. Finally a summary is given in Sec. IV.

\section{Relativistic mean field with the new interaction--FSUGold}\label{intro}\noindent
Nowadays the relativistic mean field (RMF) theory as a
density-functional approach has become a very useful tool in nuclear
physics \cite{RMF}. In the RMF theory of nuclear matter that made of
nucleons (p,n) and leptons (e, $\mu$) in a uniform magnetic field
$B$, the total interacting Lagrangian density is given by
\begin{eqnarray}
\mathcal{L} &=&\overline{\psi }_{b}(i\gamma ^{\mu }\partial _{\mu
}-M-g_{\sigma }\sigma -\frac{g_{\rho }}{2}\gamma ^{\mu }{\bm\tau }\cdot {\bm%
\rho _{\mu }}+g_{\omega }\gamma ^{\mu }\omega _{\mu }  \notag \\
&&{-q}_{{b}}{\gamma ^{\mu }\frac{1+\tau _{3}}{2}A_{\mu
}-}\frac{1}{4}\kappa _{b}\sigma _{\mu \nu }F^{\mu \nu })\psi
_{b}-\frac{1}{4}F_{\mu \nu }F^{\mu
\nu }  \notag \\
&&+\frac{1}{2}\partial _{\mu }\sigma \partial ^{\mu }\sigma -(\frac{1}{2}%
m_{\sigma }^{2}\sigma ^{2}+\frac{1}{3}g_{2}\sigma ^{3}+\frac{1}{4}%
g_{3}\sigma ^{4})  \notag \\
&&-\frac{1}{4}\Omega _{\mu \nu }\Omega ^{\mu \nu
}+\frac{1}{2}m_{\omega }^{2}\omega _{\mu }\omega ^{\mu }+\frac{\zeta
}{4!}g_{\omega }^{4}(\omega
_{\mu }\omega ^{\mu })^{2}  \notag \\
&&-\frac{1}{4}{\bm R}_{\mu \nu }\cdot {\bm R}^{\mu \nu
}+\frac{1}{2}m_{\rho }^{2}{\bm\rho }_{\mu }\cdot {\bm\rho }^{\mu
}+\Lambda _{v}g_{\rho }^{2}{\rho }_{\mu }\cdot {\rho }^{\mu
}g_{\omega }^{2}\omega _{\mu }\omega ^{\mu }
\notag \\
&&+\overline{\psi }_{l}(i\gamma ^{\mu }\partial _{\mu }-m_{l}-{q}_{{l}}{%
\gamma ^{\mu }A_{\mu }-}\frac{1}{4}\kappa _{l}\sigma _{\mu \nu
}F^{\mu \nu })\psi _{l}
\end{eqnarray}
with $A^{\mu }=(0,0,Bx,0)$ and $\sigma ^{\mu \nu }=\frac{i}{2}\left[
\gamma ^{\mu },\gamma ^{\nu }\right] $. $\kappa _{p}=1.7928\mu
_{N}$, $\kappa _{n}=-1.9130\mu _{N}$, $\kappa _{e}=1.15965\times
10^{-3}\mu _{B}$ and $\kappa _{\mu}=1.16592\times 10^{-3}\mu _{B}$
are the AMM for protons, neutrons, electrons and muons,
respectively~\cite{mass}, where $\mu _{N}$ ($\mu _{B}$) denotes the
nuclear (Bohr) magneton of nucleons (leptons). $M$, $m_{\sigma }$,
$m_{\omega }$ and $m_{\rho }$ are the nucleon-, the $\sigma $-, the
$\omega$- and the $\rho $-meson masses, respectively. The nucleon
field $\psi_{b} $ interacts with the $\sigma ,\omega ,\rho $ meson
fields $\sigma ,\omega _{\mu },\rho _{\mu }$ and with the photon
field $A_{\mu }$. The field tensors for the vector meson are given
as $\Omega _{\mu\nu }=\partial _{\mu }\omega _{\nu }-\partial _{\nu
}\omega _{\mu }$ and by similar expression for $\rho $ meson and the
photon. The self-coupling terms with coupling constants $g_{2}$ and
$g_{3}$ for the $\sigma $ meson turned out to be crucial~\cite{BB}
are introduced. Compared with the previous RMF models, the RMF
interactions employed in this work are FSUGold where two additional
parameters $\zeta$ and $\Lambda _{v}$ have been introduced: $\omega$
meson self-interactions as described by $\zeta $ which soften the
equation of state at high density, and the nonlinear mixed
isoscalar-isovector coupling described by $\Lambda _{v}$ that
modifies the density-dependence of the symmetry energy. The FSUGold
interaction gives a good description of ground state properties as
well as excitations of finite nuclei \cite{FC}. In our previous
work, this new interaction was used to study the properties of dense
matter and symmetry energy in strong magnetic fields \cite{Dong3}.

The energy spectra of the proton, neutron, electron and muon are
given by
\begin{eqnarray}
E_{\nu ,s}^{p} &=&\sqrt{k_{z}^{2}+\left( \sqrt{M^{\ast 2}+2Be\nu
}-s\kappa
_{p}B\right) ^{2}}+g_{\omega }\omega _{0}+g_{\rho }\rho _{30}, \\
E_{s}^{n} &=&\sqrt{k_{z}^{2}+\left( \sqrt{M^{\ast 2}+k_{x}^{2}+k_{y}^{2}}%
-s\kappa _{n}B\right) ^{2}}+g_{\omega }\omega _{0}-g_{\rho }\rho _{30}, \\
E_{\nu ,s}^{e} &=&\sqrt{k_{z}^{2}+\left( \sqrt{m_{e}^{2}+2Be\nu
}-s\kappa
_{e}B\right) ^{2}}, \\
E_{\nu ,s}^{\mu } &=&\sqrt{k_{z}^{2}+\left( \sqrt{m_{\mu }^{2}+2Be\nu }%
-s\kappa _{\mu }B\right) ^{2}},
\end{eqnarray}
where $\nu=0,1,2,3...$ denotes the Landau levels for charged
particles and $s = 1(-1)$ is spin-up (spin-down). The chemical
potentials $\mu$ are obtained by replacing the $k_{z}$ by $k_{f,\nu
,s}$, where $k_{z}$ is the momentum along the $z$-axis and $k_{f,\nu
,s}$ is the Fermi momentum.

\section{Magnetization of neutron star matter}\label{model}\noindent
The thermodynamical potential for the charged particle is given by
\begin{equation}
\Omega =-\frac{eB}{2\pi ^{2}}\underset{\nu ,s}{\sum }\int_{0}^{\infty }dk_{z}%
\frac{1}{\beta }\ln \left[ 1+e^{-\beta (E_{\nu ,s}-\mu )}\right] .
\end{equation}
where the contribution of antiparticles is not taken into account.
The magnetization $M=-\left( \frac{\partial \Omega }{\partial
B}\right) _{T,V,\mu }$ takes the form
\begin{eqnarray}
M &=&\frac{e}{2\pi ^{2}}\underset{\nu ,s}{\sum }\int_{0}^{\infty }dk_{z}%
\frac{1}{\beta }\ln \left[ 1+e^{-\beta (E_{\nu ,s}-\mu )}\right]   \notag \\
&&-\frac{eB}{2\pi ^{2}}\underset{\nu ,s}{\sum }\int_{0}^{\infty }dk_{z}\frac{%
e^{-\beta (E_{\nu ,s}-\mu )}}{1+e^{-\beta (E_{\nu ,s}-\mu
)}}\frac{\partial E_{\nu ,s}}{\partial B}.
\end{eqnarray}
In the zero-temperature limit the proton magnetization is
\begin{eqnarray}
M_{p} &=&-\frac{\varepsilon _{p}}{B}+\frac{\rho _{p}E_{f}^{p}}{B}-\frac{eB}{%
2\pi ^{2}}\underset{\nu ,s}{\sum }\left( \sqrt{M^{\ast 2}+2e\nu
B}-s\kappa
_{p}B\right) \times  \nonumber \\
&&\left( \frac{e\nu }{\sqrt{M^{\ast 2}+2e\nu B}}-s\kappa _{p}\right)
\ln \left\vert \frac{E_{f}^{p}+k_{f,\nu ,s}^{p}}{\sqrt{M^{\ast
2}+2Be\nu }-s\kappa _{p}B}\right\vert,\label{A}
\end{eqnarray}
where $\rho_{p}$ is the proton density and the energy density of the
proton is
\begin{eqnarray}
\varepsilon _{p}&=&\frac{eB}{4\pi ^{2}}\underset{\nu ,s}{\sum }\bigg
[ k_{f,\nu ,s}^{p}E_{f}^{p}+\left( \sqrt{M^{\ast 2}+2\nu eB}-s\kappa
_{p}B\right)^{2} \nonumber \\
&& \times \ln \left\vert \frac{k_{f,\nu ,s}^{p}+E_{f}^{p}}{\sqrt{M^{\ast 2}+2\nu eB%
}-s\kappa _{p}B}\right\vert \bigg].
\end{eqnarray}
Here the feeble change of $\sigma$-field is neglected. Similar
expression can be obtained for the electron and muon. For
simplicity, we calculate the neutron magnetization with $M_{n}=(\rho
_{n\uparrow }-\rho _{n\downarrow })\kappa _{n}$. The magnetic
susceptibility is given as $\chi =M/B$. We would like to stress
that, due to their Landau diamagnetism, there are not such a simple
relation between the magnetization and spin polarization for the
charged particles.

\begin{figure*}[htbp]
\begin{center}
\includegraphics[width=1.0\textwidth]{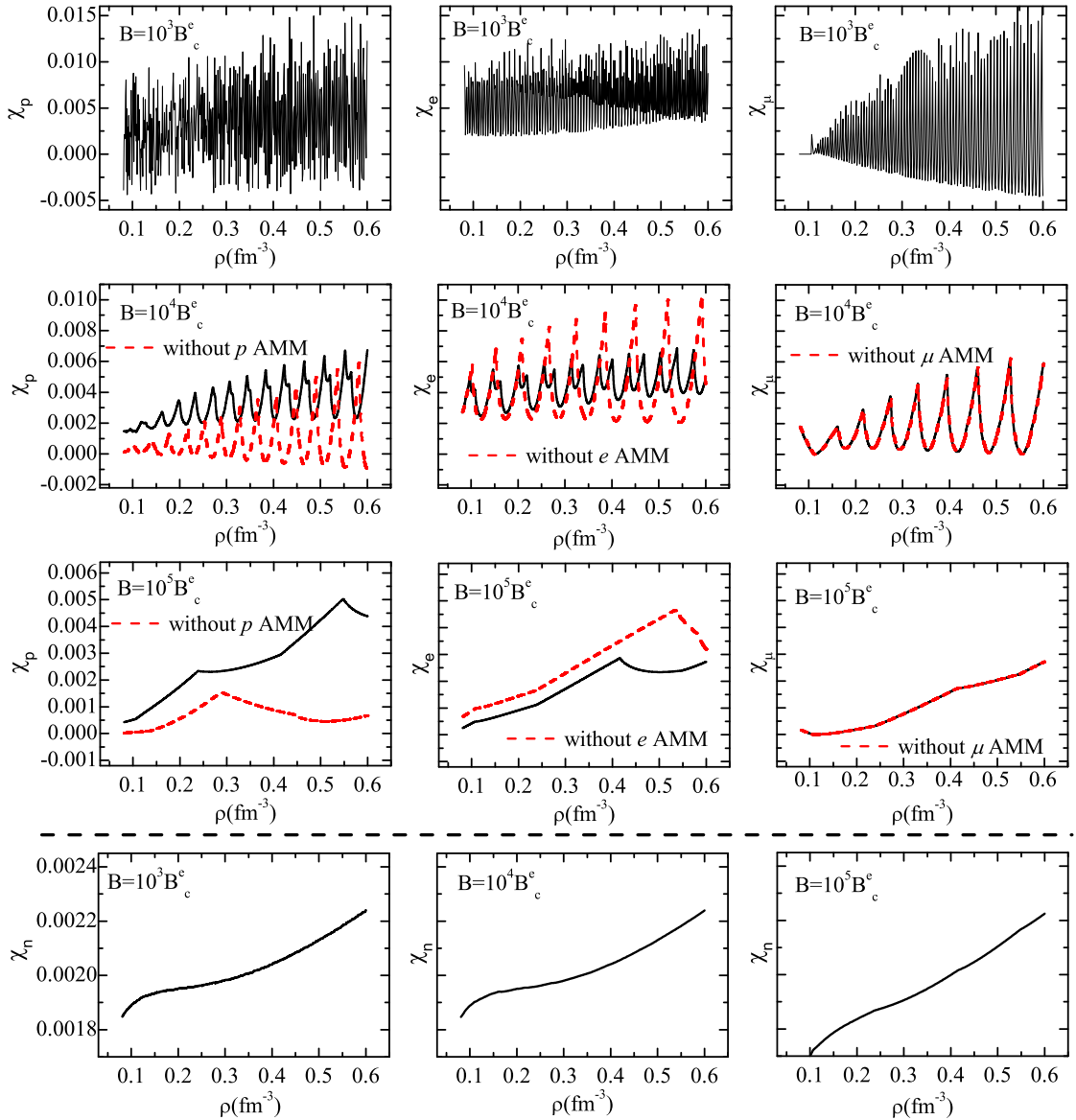}
\caption{(Color online) Magnetic susceptibilities of $n$, $p$, $e$
and $\mu$ for the $\beta$-stable matter as a function of nuclear
matter density for different values of the magnetic field $B$. The
subgraphs above the dashed line are magnetic susceptibilities for
charged particles. The neutron magnetic susceptibilities are
displayed in the last row. The magnetic field is in unit of the
electron critical field $B_{c}^{e}=4.414\times10^{13}$G.}
\end{center}
\end{figure*}

\begin{figure*}[htbp]
\begin{center}
\includegraphics[width=0.9\textwidth]{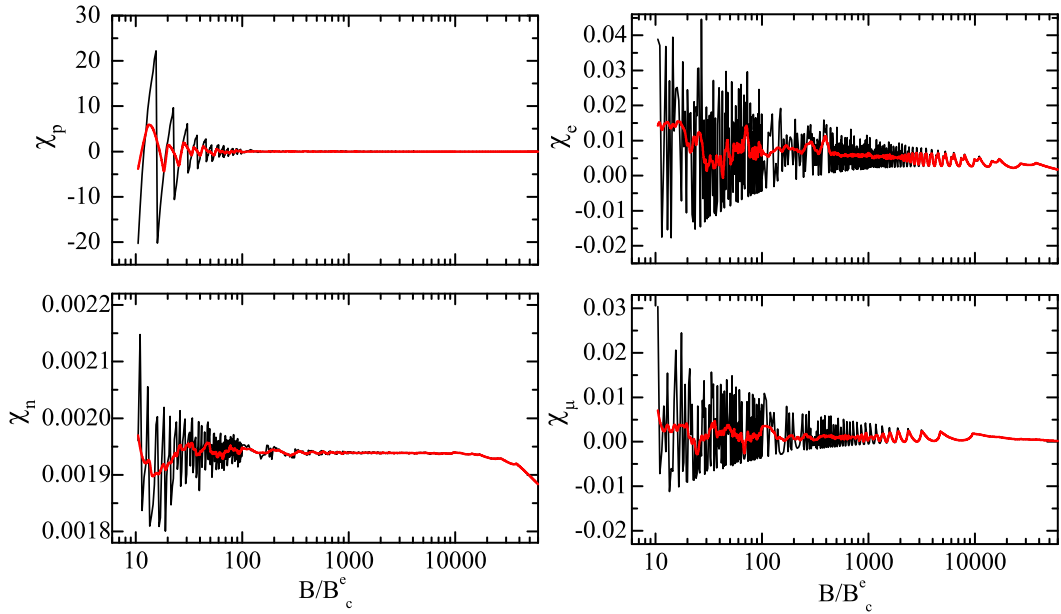}
\caption{Magnetic susceptibilities of $n$, $p$, $e$ and $\mu$ for
the $\beta$-stable matter as a function of the magnetic field
strength $B$. The red bold curves correspond to average over a long
period. The density we selected is $\rho=0.16$ fm$^{-3}$ as an
example.}
\end{center}
\end{figure*}

The magnetic susceptibilities of the proton, electron, muon and
neutron versus the density $\rho$ for the $\beta$-stable neutron
star matter under different magnetic fields are presented in Fig. 1.
For the charged particles, the magnetic susceptibilities show the
oscillations in particular in the case of a relative weak magnetic
field. Besides, as shown in the top three panels of Fig. 1, they are
positive in most cases and sometimes fall into their negative ranges
in the case of rather weak magnetic fields. The 'oscillation period'
of the magnetic susceptibility depends on the density of the Landau
energy states. This density of state reduces with the increase of
the magnetic field strength, and hence the 'oscillation period'. It
is found that the magnetic susceptibilities of the charged particles
tend to be larger than that of the neutron, indicating that the
neutron star matter can not be treated simply as the pure neutron
matter when one studies its magnetization. Neutrons carry no charge
so that they have no Landau levels to fill. Hence, the direct
coupling of neutrons to magnetic field is just due to the neutron
AMM. For the protons, electrons and muons, however, their charge
strongly couples with the magnetic field forming the Landau levels,
and this coupling is much stronger than the direct coupling between
the AMM and magnetic field. Roughly speaking, the more the Landau
levels are, the stronger the magnetization. In an extreme case that
the particles occupy the Landau ground state (only one Landau
level), the magnetization vanishes due to the Landau diamagnetism
being counterbalanced by the Pauli paramagnetism if one ignores the
AMM. Our calculations indicate that the magnetization of the
electrons is only a few percent, which is not much larger than these
of other components. Accordingly, in contradiction with the
investigation of Ref. \cite{Peng}, the primal magnetic field of the
neutron stars can not be greatly boosted up by the magnetization of
the highly degenerate relativistic electron gas. The fundamental
reason is that the Pauli paramagnetism is canceled out to a large
degree by the diamagnetism for the electron.

\begin{figure}[htbp]
\begin{center}
\includegraphics[width=0.5\textwidth]{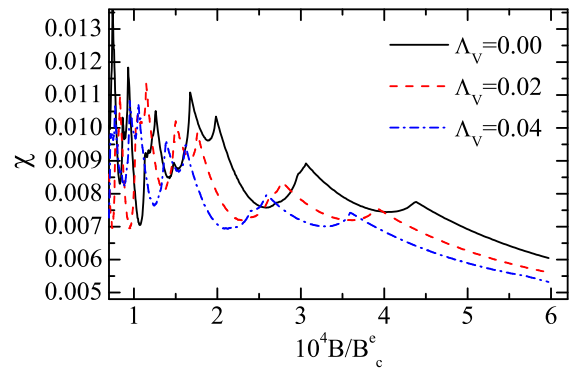}
\caption{(Color online) Magnetic susceptibilities of the neutron
star matter versus the magnetic field strength $B$. The density we
selected is $\rho=0.16$ fm$^{-3}$ as an example. The calculations
are performed with the modified FSUGold interactions \cite{BKSP}
providing stiff ($\Lambda _{v}=0.00$) to soft ($\Lambda _{v}=0.04$)
symmetry energy.}
\end{center}
\end{figure}

To show the effects of the AMM of each component on the magnetic
susceptibility $\chi$, we present the calculated $\chi_{p}$,
$\chi_{e}$, $\chi_{\mu}$ without the inclusion of the their AMM in
the middle six panels of Fig. 1 remarked by dash curves for
comparison. The effect of the muon AMM can be neglected completely
because of its quite small value (about 1/207 of electron AMM). The
proton and electron AMM affect their respective magnetic
susceptibility evidently. With the inclusion of the AMM, the doubly
degeneracy with opposite spin projections is destroyed and hence the
peaks and shapes of the curves are modified. On the whole, the
proton AMM leads to an enhancement of $\chi_{p}$ while the electron
AMM causes the $\chi_{e}$ reduce slightly, which has connection with
the spin polarization--the positive polarizability for protons but
negative one for electrons. Compared with the proton AMM, the effect
of the electron AMM is weaker because the electron AMM is about
thousandth of its normal magnetic moment while the proton AMM shares
the same order of magnitude as its normal magnetic moment.

The magnetic susceptibility versus the magnetic field strength are
presented in Fig. 2 taking the $\beta$-stable matter at $\rho=0.16$
fm$^{-3}$ as an example. The detailed structure of the magnetization
exhibits strong de Haas--van Alphen oscillations. The amplitudes of
the oscillations become increasing small as the magnetic field
strength increases, and the magnetic susceptibilities of the charged
particles tend toward zero when the magnetic field is very strong.
The reason lies in the reduction of the Landau levels as the
magnetic field strength increases. Though the neutron has no Landau
levels to occupy, its magnetic susceptibility also fluctuates with
the magnetic field owing to the fact that the magnetic field affects
the neutron density at a given nucleon density. One conspicuous
phenomenon is that the absolute value of the proton magnetization
$M_{p}$ tends to be much larger than the $M_{e}$, $M_{\mu}$ and
$M_{n}$ in a relative weak magnetic field--that is, the proton is
much stronger magnetized compared with other components. One can
easily realize from the relevant discussions about the Fig. 1. When
the magnetic field is weak, the induced magnetic field due to the
magnetization can be much stronger than the original field but
fluctuated wildly. The irregularity oscillations can be averaged to
smooth out the wild oscillations to a large extent, being analogous
to the averaged viscosities in the presence of strong magnetic
fields that discussed in Ref. \cite{XU0}. The strong magnetization
perhaps has something to do with the origin of the magnetic field in
neutron stars: The original seed field is gradually amplified by the
magnetization. Of course, it needs further investigation.

The density-dependent symmetry energy plays a crucial role in
understanding a variety of issues in nuclear physics as well as
astrophysics \cite{PD,AWS,VB,BAL,JML,NS1,BKSP,JD,DONG0}. Fig. 3
displays the total magnetic susceptibility $\chi$ as a function of
the magnetic field strength with the modified FSUGold interactions
which yield stiff to soft symmetry energy, where $\Lambda_{v}$ is
varied while $g_{\rho}$ is adjusted so that for each $\Lambda_{v}$
the asymmetry energy remains fixed at a given density and this
prescription ensures that the binding energy as well as the proton
density of a heavy nucleus, such as $^{208}$Pb, are within the
measured values \cite{BKSP}. The interaction with a stiff symmetry
energy tends to yield a large $\chi$ at a strong magnetic field
$B>10^{4}B_{c}^{e}$ and the 'peaks' shift forward compared with that
yields a soft symmetry energy. These stem from the fact that a
stiffer symmetry energy gives a lower neutron fraction. As a
consequence, the effects of the symmetry energy on the magnetization
are distinct at strong magnetic fields.

\section{Summary}\label{sec5}\noindent

The magnetization of neutron star matter in magnetic fields has been
studied within the FSUGold interaction. The present analysis is
based on the zero-temperature limit for simplicity since the Fermi
temperature is much larger than the real temperature in normal
neutron stars. The main conclusion are summarized as follows. (1)
The magnetic susceptibilities of the neutron is not dominant,
indicating the neutron star matter can not be treated as the pure
neutron matter for simplicity when one studies its magnetization.
(2) Being inconsistent with the conclusion in Ref. \cite{Peng}, the
small electron magnetic susceptibility indicates the observed
super-strong magnetic field of neutron stars does not originate from
the induced Pauli paramagnetism of the highly degenerate
relativistic electron gas in the neutron star interiors. (3) The
proton and electron AMM affect their respective magnetic
susceptibility evidently whereas the muon AMM can be neglected
completely. The role of the AMM of the neutron, proton and electron
suggested they can not be discarded arbitrarily. (4) The proton is
found to be much stronger magnetized compared with other components
when the magnetic field is relatively weak ($B < 10^{2}B_{c}^{e}$).
The magnetization of the matter can be appear to be very large,
which differs from the conclusion in Ref. \cite{GM3}. The
calculation in Ref. \cite{GM3} was correct, but it did not include
the case of the very low magnetic fields so that it concluded the
magnetic susceptibility is only a few percent. The magnetization
perhaps is related to the origin of the strong magnetic field in
neutron stars, but it needs to be explored further. (5) The
magnetization of neutron star matter is affected distinctly by the
density-dependent symmetry energy.

At low temperature and weak fields, pairing correlations may
dominate the magnetic susceptibility. Pairing in the $^{1}S_{0}$ and
$^{3}PF_{2}$ channels may have a large impact on the magnetic
response of the system, which needs to be further investigated.

\section*{Acknowledgements}
Dong is thankful to Long-Jun Wang for providing useful help. This
work was supported by the 973 Program of China(No. 2013CB834405), the National Natural
Science Foundation of China under Grants No. 11175219, 10975190,
11275271; the Knowledge Innovation Project (KJCX2-EW-N01) of Chinese
Academy of Sciences, CAS/SAFEA International Partnership Program for
Creative Research Teams (CXTD-J2005-1)and the Funds for Creative
Research Groups of China under Grant No. 11021504.

\end{CJK*}
\end{document}